\documentclass{article}
\usepackage{iclr2027_conference}
\iclrfinalcopy
\usepackage{times}
\usepackage{amsmath,amssymb}
\usepackage{graphicx}
\usepackage{float}
\usepackage{wrapfig}
\usepackage{flafter}
\usepackage{booktabs}
\usepackage{xcolor,colortbl}
\usepackage{enumitem}
\usepackage{algorithm}
\usepackage{algpseudocode}
\usepackage{xspace}
\usepackage[hidelinks]{hyperref}
\usepackage{url}

\makeatletter
\newcommand{\figurecaption}{\def\@captype{figure}\caption}
\makeatother

\newcommand{\cmark}{\checkmark}
\newcommand{\xmark}{\textcolor{gray}{--}}
\newcommand{\agepa}{Adaptive-\textsc{Gepa}\xspace}
\newcommand{\gepa}{\textsc{Gepa}\xspace}
\newcommand{\fpa}{\textsc{Gepa-Fpa}\xspace}
\newcommand{\pnull}{\ensuremath{P_0}\xspace}

\newcommand{\dfeed}{\mathcal{D}_{\mathrm{feedback}}}
\newcommand{\dpareto}{\mathcal{D}_{\mathrm{val}}}

\title{\agepa: Make Your Harness Fit\\Heterogeneous Requests}
\author{%
\textbf{Tianyu Chen}\textsuperscript{1,}\thanks{Equal contribution.}\quad
\textbf{Yasi Zhang}\textsuperscript{2,}\footnotemark[1]\quad
\textbf{Ruiyi Wang}\textsuperscript{3}\\
\textbf{Xinran Zhao}\textsuperscript{4}\quad
\textbf{Taoran Li}\textsuperscript{5}\quad
\textbf{Mingyuan Zhou}\textsuperscript{6}\\[0.5em]
\normalfont\small\textsuperscript{1}\,University of Texas at Austin\quad
\normalfont\small\textsuperscript{2}\,University of California, Los Angeles\\
\normalfont\small\textsuperscript{3}\,University of California, San Diego\quad
\normalfont\small\textsuperscript{4}\,Carnegie Mellon University\\
\normalfont\small\textsuperscript{5}\,Shenzhen Institutes of Advanced Technology\quad
\normalfont\small\textsuperscript{6}\,Microsoft}
\hypersetup{%
  pdftitle={Adaptive-GEPA: Make Your Harness Fit Heterogeneous Requests},
  pdfauthor={Tianyu Chen; Yasi Zhang; Ruiyi Wang; Xinran Zhao; Taoran Li; Mingyuan Zhou},
  pdfsubject={Preprint}}
\begin{document}
\raggedbottom
\maketitle
\lhead{Preprint}
\begin{abstract}

Reflective optimizers such as \gepa \citep{gepa2025} improve language-model prompts from execution traces and evaluator feedback; full-program extensions can also rewrite tools and control flow.
In practice, a user hands the same endpoint heterogeneous requests whose effective solutions require different tools, reasoning modes, and control flow.
Optimizing one shared program leaves this division of work implicit in source-code search, while optimizing a separate program per request family fixes it beforehand.
We introduce \agepa, which learns both how to divide requests and how to solve them.
It evolves a router and a library of specialist programs under one search budget.
The router's instructions, each specialist's description, and its program code are plain, human-readable text, edited from feedback.
To combine branches, it aligns specialists by the requests they handle and inherits descriptions together with programs.
On a fixed mixture of four task families, the reported Qwen3-8B run evolves four experts without supplying family labels to the router or reflection model; its routing matches the task partition on all 651 test requests.
Its family-mean test score ($\times100$) rises from $52.6$ to $70.6$, compared with $62.5$ for \gepa's full-program adapter and $54.0$ for GRPO at a nominal budget of 18,000 scored calls.
These counts do not equate total compute.
Figure~\ref{fig:main-story} summarizes the learning curves, final test scores, and routing agreement.

\end{abstract}

\begin{figure}[!ht]
\centering
\includegraphics[width=\linewidth,trim=0 0.15in 0 0.08in,clip]{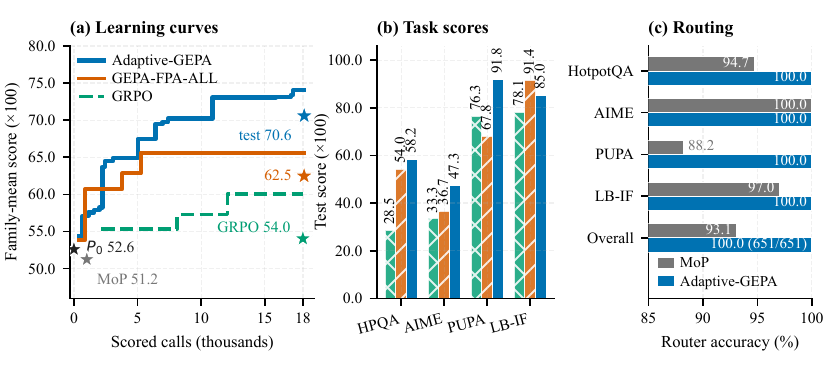}
\caption{\textbf{One routed search improves a mixed workload.} (a) Best validation means; stars mark test scores. (b) Per-task test scores. (c) Post-hoc family agreement ($85$--$100\%$ axis). Nominal budgets are 18,000 scored calls for search and GRPO, and 1,010 for MoP.}
\label{fig:main-story}
\label{fig:budget}
\end{figure}
\clearpage

\begin{figure}[H]
\centering
\includegraphics[width=\linewidth]{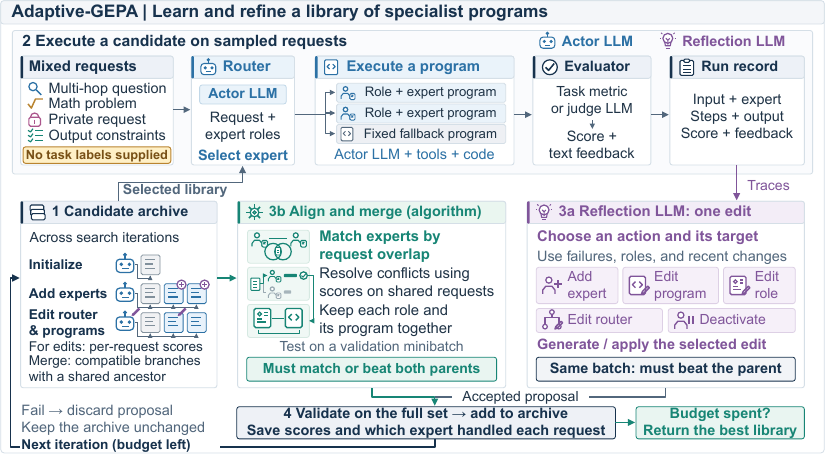}
\caption{\textbf{\agepa learns both the division of work and the programs that carry it out.} The actor LLM (blue) routes requests and supplies model calls inside programs. The reflection LLM (purple) selects actions and generates any needed code or text edits. The search algorithm matches experts by request overlap, resolves conflicts using scores on shared requests, and retains each role--program pair intact. The schematic score bars and check illustrate expert selection. Passing proposals receive full validation and enter the archive. The snapshots show growth and edits across iterations; gray cards denote the fixed fallback, plus signs additions, and pencils edits. \textbf{No task labels are supplied to the router or reflection model}.}
\label{fig:search-overview}
\end{figure}

\section{Introduction}
\label{sec:intro}

A shared language-model service receives requests that need different procedures, for example, factual retrieval, mathematical reasoning, privacy protection, or constrained writing.
Improving such a service requires learning both useful procedures and when each procedure should be used.

Program optimization can revise instructions, tool calls, and control flow using execution feedback~\citep{dspy2024,adas2024,aflow2025,gepa2025}. One option is to optimize a single program over all requests. Another is to optimize each known task separately and supply a dispatcher. The first leaves the division of work implicit in source-code search; the second fixes it beforehand. A single program can also implement dispatch: the distinction concerns how search organizes and changes the system, rather than what code can express.

We study a third option: \emph{learn the division of work and the programs together in one search}. Search must decide when to create a specialist, what it should handle, and how to improve it. It must also combine specialists from different branches. Slot numbers cannot reliably identify their roles: the same capability can occupy different slots, while the same slot can serve different capabilities.

\agepa addresses these choices by extending the reflective search of \gepa~\citep{gepa2025} to a routed program library (Figure~\ref{fig:search-overview}). Each specialist pairs a short description of its responsibility with executable source code. A reflection model first identifies the needed change, then edits the selected component. The descriptions let it reason about the other specialists without reading all their source code. To merge two libraries, the method matches specialists using the validation requests they actually handled, then keeps each description together with its program. This makes the division of work explicit throughout search, from local edits to branch combination.

We evaluate a fixed mixture of HotpotQA, AIME, PUPA, and LiveBench instruction following. Labels support sampling and scoring but supply no target expert assignment to the models. At approximately 18,000 scored calls, the Qwen3-8B library reaches $70.6$, versus $62.5$ for single-program evolution and $73.0$ for independently optimized, label-routed programs. Its four specialists learn different retrieval, reasoning, verification, and repair procedures.

Our contribution is an editable router--description--program representation and a request-based rule for inheriting specialist pairs across branches. Joint optimization and heterogeneous routing also appear in recent systems (Section~\ref{sec:related}); we focus on how specialist identity is maintained during search and recombination. Our experiments compare complete systems on this fixed mixture, without establishing seed robustness or isolating component effects.

\section{Problem Setting}
\label{sec:problem}

A system serving mixed requests may need retrieval for one request, extended reasoning for another, and a short transformation for a third.
Optimizing such a system requires deciding both what each program should do and which requests it should handle.

\paragraph{Compound systems and full-program optimization.}
Following \citet{gepa2025}, a compound AI system combines language-model modules with control flow and tool calls.
We write its specification and execution as
\begin{equation}
    M=(\langle m_1,\ldots,m_J\rangle,c),
    \qquad m_i=(\pi_i,\theta_i),
    \qquad \Phi_M:\mathcal X\rightarrow\mathcal Y,
    \label{eq:compound-system}
\end{equation}
where $\pi_i$ is a module's prompt, $\theta_i$ its model weights, and $c$ the control flow connecting modules and tools.
Input and output schemas are part of the specification and omitted from the notation.
Prompt optimization changes $\Pi_M=\langle\pi_1,\ldots,\pi_J\rangle$ within a fixed program.
Full-program optimization also changes the modules, control flow, tool use, and inference configuration.
Here, $M$ is represented as executable source, and all underlying model weights remain fixed.

\paragraph{Mixed requests and objective.}
Let $\mathcal D$ be a mixture of $L$ task distributions $T_1,\ldots,T_L$.
An incoming request includes task instructions, but no task-family label is supplied to the system.
We optimize a routed program $P=(R,\{M_k\}_{k=1}^{K})$, where $R$ chooses a specialist and each $M_k$ is a complete program of the form above.
Both the programs and their division of requests are learned; $K$ need not equal $L$.
An evaluator scores output $y=\Phi_P(x)$ with a scalar $\mu(x,y)\in\mathbb R$ and may return textual feedback $\phi(x,y)$.
Reference answers and other evaluator metadata are implicit in these functions.
The objective is
\begin{equation}
    P^\star\in\arg\max_P\;
    \mathbb E_{x\sim\mathcal D}\!\left[\mu\bigl(x,\Phi_P(x)\bigr)\right],
    \qquad N_{\mathrm{eval}}<B\text{ at iteration start}.
    \label{eq:agepa-objective}
\end{equation}
Here $N_{\mathrm{eval}}$ counts scored executions: each runs the system on one request and evaluates its output. Search starts another iteration only while $N_{\mathrm{eval}}<B$, and finishes any evaluations in that iteration; the final count can therefore exceed the nominal budget.
The budget constrains the search used to find $P^\star$, rather than the internal model or tool calls in a deployed program.

\section{Adaptive-GEPA}
\label{sec:method}

\agepa searches over both request allocation and executable behavior.
It edits the router, each specialist's responsibility description and source code, and which specialists are active (Table~\ref{tab:search-design}).
Full-program search can also generate internal dispatch; our design makes these choices separate search targets.
A local edit develops one part of a candidate, while an aligned merge combines specialists from different branches.
Algorithm~\ref{alg:agepa} summarizes the search.

\begin{table}[t]
\centering
\caption{\textbf{\agepa makes routing, responsibilities, and programs separate search targets.} FPA denotes GEPA's full-program adapter; MoP is Mixture-of-Prompts. Original GEPA is a design reference; the other rows describe the evaluated configurations.}
\label{tab:search-design}
\small
\setlength{\tabcolsep}{3pt}
\renewcommand{\arraystretch}{1.12}
\newcommand{\designcell}[2]{\parbox[t]{#1\linewidth}{\raggedright\strut#2\strut}}
\begin{tabular}{@{}llll@{}}
\toprule
\designcell{0.21}{Method} & \designcell{0.27}{Search variables} & \designcell{0.22}{Request allocation} & \designcell{0.25}{Update organization} \\
\midrule
\designcell{0.21}{Original \gepa~\citep{gepa2025}} & \designcell{0.27}{Module prompts} & \designcell{0.22}{Supplied workflow; prompts editable} & \designcell{0.25}{Round-robin edits; optional module merges} \\
\addlinespace[3pt]
\designcell{0.21}{MoP~\citep{mop2024}} & \designcell{0.27}{Cluster count, instructions, demo assignments} & \designcell{0.22}{Nearest embedding centroid} & \designcell{0.25}{Clustering, then prompt assignment} \\
\addlinespace[3pt]
\designcell{0.21}{\fpa-ALL} & \designcell{0.27}{One complete program} & \designcell{0.22}{Dispatch can be generated in source} & \designcell{0.25}{Full-program edits} \\
\addlinespace[3pt]
\designcell{0.21}{Per-family \fpa} & \designcell{0.27}{A program per known family} & \designcell{0.22}{True task labels} & \designcell{0.25}{Separate searches} \\
\addlinespace[3pt]
\rowcolor{blue!7}
\designcell{0.21}{\agepa} & \designcell{0.27}{Router, descriptions, programs, active slots} & \designcell{0.22}{Router reads requests and descriptions} & \designcell{0.25}{Feedback selects action and target; aligned merges} \\
\bottomrule
\end{tabular}
\end{table}

\subsection{Representing the division of work}

Each specialist has a natural-language description $d_k$, a program $M_k$, and an activity indicator $z_k\in\{0,1\}$.
The full candidate and its active slots are
\begin{equation}
    P=\bigl(R,\{(d_k,M_k,z_k)\}_{k=1}^{K_{\max}}\bigr),
    \qquad
    \mathcal K(P)=\{k\in\{1,\ldots,K_{\max}\}:z_k=1\}.
    \label{eq:candidate-dispatch}
\end{equation}
The prompted language-model router reads the request and active descriptions, then dispatches it as
\begin{equation}
    \hat k=R\bigl(x;\{(k,d_k):k\in\mathcal K(P)\}\bigr)
        \in\mathcal K(P)\cup\{0\},
    \qquad
    \Phi_P(x)=\Phi_{M_{e_P(x)}}(x).
    \label{eq:routed-execution}
\end{equation}
Here $M_0=\pnull$ is the fixed generic fallback, and $e_P(x)$ is the slot that actually executes after invalid or unavailable selections resolve to fallback.
All specialist slots start inactive; search changes their number $K=|\mathcal K(P)|$ up to $K_{\max}$, set to five in the main experiments.
The router sees descriptions, not program source.
A specialist can execute multiple model calls and tools.

Search uses a feedback set $\dfeed$ for reflection and a separate validation set $\dpareto$ for selection.
For any evaluation set $S$, define the measured mean score
\begin{equation}
    \bar\mu_S(P)=\frac{1}{|S|}\sum_{x\in S}\mu\bigl(x,\Phi_P(x)\bigr).
    \label{eq:empirical-score}
\end{equation}
These means refer to the recorded evaluations; separate candidates may have identical source and different sampled scores.
Task-family labels are withheld from the router, specialists, and reflection model.
Labels support sampling and evaluator selection, but supply no target partition.

\subsection{Choose an action, then edit its target}
\label{sec:method:actions}

Router and description edits can redirect requests, but cannot repair an existing program's code.
Program edits address faulty procedures; activation supplies missing ones.
Section~\ref{sec:results:edit-allocation} illustrates this distinction: a program rewrite restores a specialist without changing the router or descriptions.

A description states one specialist's coverage; the router resolves competing responsibilities.
In Section~\ref{sec:results:library}, a broader retrieval description attracts missed factual requests but also constrained rewrites. A router edit then prioritizes the requested operation over topic words.
These choices could share a prompt; separate targets let search revise either one expert's coverage or the common selection rule.

The appropriate edit depends on the failure, so reflection first selects an action $a$ and, when needed, a slot $k$.
It uses the router, descriptions, usage, action history, and minibatch records $\tau=(x,\hat k,t,y,\mu(x,y),\phi(x,y))$, where $t$ contains execution steps and errors; it sees no specialist source.
A second call writes the edit, reading only the target source for program or description rewrites.
Actions are selected from feedback, not stages in a fixed cycle.

Activation writes $(d_k,M_k)$ and sets $z_k=1$; rewrites change only $d_k$, $M_k$, or $R$; deactivation sets $z_k=0$ without another call.
Appendix~\ref{app:actions} specifies access; unedited specialist source stays fixed.

\subsection{Align specialists before merging branches}
\label{sec:method:merge}

Different branches can assign the same slot index to unrelated roles, so copying by index can overwrite the wrong specialist.
We align specialists using handled validation requests, without task labels.
Section~\ref{sec:results:evolution} illustrates this mismatch and its alignment.

For validation inputs $x_1,\ldots,x_N$, executed slots $e_P(x_n)$ record which specialist handled each request.
Define specialist $i$'s \emph{footprint} and its Jaccard overlap with specialist $j$ in parent $P'$ as
\begin{equation}
    F_i(P)=\{n:e_P(x_n)=i\},
    \qquad
    J_{ij}(P,P')=
    \frac{|F_i(P)\cap F_j(P')|}{|F_i(P)\cup F_j(P')|},
    \label{eq:footprint-jaccard}
\end{equation}
with zero overlap for an empty union.
The initial alignment uses mutual best matches:
\begin{equation}
    j=\arg\max_{j'}J_{ij'},\qquad
    i=\arg\max_{i'}J_{i'j},\qquad J_{ij}>0,
    \label{eq:mutual-footprint-matching}
\end{equation}
with deterministic tie breaking.
This is a local matching heuristic over executed traffic, not a maximum-weight assignment or a test of semantic equivalence.
Unmatched specialists first use inactive slots.
Appendix~\ref{app:method-details} specifies the capacity fallback for other positive-overlap matches.

To inherit each program with its stated role, a merge copies $(d_k,M_k)$ together.
If both parents changed an aligned pair, the higher archived mean on $F_i(P)\cap F_j(P')$ determines which pair to keep.
This compares specialists on requests they both handled; Appendix~\ref{app:method-details} gives missing-evidence rules.
Alignment and selection reuse validation records, but the assembled child requires a new evaluation because the combined library can change request allocation.

\subsection{Search, acceptance, and final selection}
\label{sec:method:training}

\begin{wrapalgorithm}{r}{0.5\linewidth}
\caption{\agepa}
\label{alg:agepa}
\footnotesize
\algrenewcommand{\algorithmicrequire}{\textbf{Input:}}
\algrenewcommand{\algorithmicindent}{0.8em}
\begin{algorithmic}[1]
\Require $M_0$, $\dfeed$, $\dpareto$, $B$
\State Validate seed; initialize $\mathcal P$
\While{$N_{\mathrm{eval}}<B$}
    \If{merge scheduled; $P,P'$ eligible}
        \State $P^+\gets\Call{AlignAndMerge}{P,P'}$
        \State Sample $S\subset\dpareto$; score $P^+$ on $S$
        \State $\mathrm{keep}\gets\bar\mu_S(P^+)\geq\max\{\bar\mu_S(P),\bar\mu_S(P')\}$
    \Else
        \State $P\gets\Call{ParetoSelect}{\mathcal P}$
        \State Sample $\mathcal B\subset\dfeed$; $\tau\gets\Call{Execute}{P,\mathcal B}$
        \State $(a,k)\gets\Call{SelectAction}{\tau}$
        \State $P^+\gets\Call{EditTarget}{P,a,k,\tau}$
        \State Score $P^+$ on $\mathcal B$
        \State $\mathrm{keep}\gets\bar\mu_{\mathcal B}(P^+)>\bar\mu_{\mathcal B}(P)$
    \EndIf
    \State \textbf{if} $\mathrm{keep}$: fully validate and archive $P^+$
\EndWhile
\State \Return $\widehat P$ (Equation~\ref{eq:final-selection})
\end{algorithmic}
\end{wrapalgorithm}

The candidate pool $\mathcal P$ stores alternative programs and their per-example validation scores.
\agepa retains \gepa's per-request Pareto selection: its \emph{frontier} contains candidates that attain the best score on at least one validation request, after removing redundant coverage.
A reflected child must strictly improve on its parent on the feedback minibatch.
A merged child instead uses a validation subsample and must match or exceed the stronger parent there.
Passing either check triggers full validation and insertion into $\mathcal P$.
We return
\begin{equation}
    \widehat P=\arg\max_{P\in\mathcal P}\bar\mu_{\dpareto}(P).
    \label{eq:final-selection}
\end{equation}
After full evaluation of a merge, specialists receiving no routing traffic are deactivated.
The implementation retains the pre-cleanup scores; Appendix~\ref{app:method-details} explains this approximation and the full search procedure.

\section{Experiments}
\label{sec:results}
\label{sec:setup}

We compare complete systems, then use recorded traces to examine how their libraries change.

\subsection{Evaluation setting}

\paragraph{Tasks and scores.}
The workload combines multi-hop factual answering (HotpotQA~\citep{hotpotqa2018}), competition mathematics (AIME), privacy-preserving delegation (PUPA~\citep{papillon2025}), and constrained text generation (LiveBench-IF~\citep{livebench2024,ifeval2023}).
Higher scores are better: HotpotQA uses token F1, AIME uses exact match, PUPA averages answer quality and one minus privacy leakage, and LiveBench-IF combines complete and partial constraint satisfaction.
Each AIME problem is attempted five times independently; we average these attempts to estimate pass@1, not pass@5.
Our primary score is the unweighted mean of the four family scores,
\begin{equation}
S_{\mathrm{mean}}(P)=\frac{1}{4}\sum_{f=1}^{4}S_f(P).
\label{eq:family-mean}
\end{equation}
This gives each capability equal weight despite different test-set sizes. We report all task scores and their means multiplied by 100. Appendix~\ref{app:data-details} describes the splits and feedback.

\paragraph{Models and search budget.}
The main execution model is Qwen3-8B~\citep{qwen32025}; GPT-5.5 supplies reflection. Programs can use ColBERTv2 retrieval~\citep{colbertv22022} and choose Qwen3's non-thinking (fast) or thinking (deep) execution profile. We allow five specialist slots and use seed 43.
The nominal budget is 18,000 \emph{scored calls}, each evaluating one program on one example. The completed \agepa and single-program runs use 18,103 and 18,096 calls because stopping follows a batch.
This budget does not equate model generations, tokens, or total compute: a program can make several model calls before receiving one score.
Selection uses validation mean alone. Researchers saw periodic test results during development (Section~\ref{sec:limitations}).

\paragraph{Comparisons.}
Our closest baseline, \textbf{\fpa-ALL}, evolves one complete program with the same mixed data, tools, reflection model, and nominal budget. Its selected program implements internal dispatch but uses one global execution profile; \agepa chooses one per specialist.
\textbf{Per-family \fpa} independently optimizes four programs at 4,500 calls each, then selects one using the true task label. It is a reference, not a strict upper bound.
We also report the \textbf{generic seed}, \textbf{Mixture-of-Prompts (MoP)}~\citep{mop2024}, and \textbf{GRPO}~\citep{grpo2024}.
MoP uses its native 1,010-call schedule. GRPO fine-tunes all Qwen3-8B weights without retrieval or reflection, at 18k and a retrospective 90k cutoff (Appendix~\ref{app:baselines}). These are contextual comparisons; they do not rank optimization paradigms under matched capabilities or compute.

\subsection{A joint search improves the mixed workload}
\label{sec:results:main}

\begin{table}[htbp]
\centering
\caption{\textbf{One joint search approaches independently optimized, label-routed programs.} Qwen3-8B test scores (seed 43). Means weight families (primary) or examples equally; $n$ is the test-set size. Bold marks the best score without task labels. Budgets count scored calls; program search uses nominal caps, and the 90k GRPO cutoff is retrospective.}
\label{tab:main_results}
\small
\setlength{\tabcolsep}{8.2pt}
\begin{tabular}{@{}>{\columncolor{white}[0pt][\tabcolsep]}lrrrrrr>{\columncolor{white}[\tabcolsep][0pt]}r@{}}
\toprule
 & Budget & \multicolumn{4}{c}{Test score ($\times100$) $\uparrow$} & \multicolumn{2}{c}{Mean ($\times100$) $\uparrow$} \\
\cmidrule(lr){3-6}\cmidrule(l){7-8}
Method & (calls) & \shortstack{HotpotQA\\{\scriptsize$n=300$}} & \shortstack{AIME\\{\scriptsize$n=30$}} & \shortstack{PUPA\\{\scriptsize$n=221$}} & \shortstack{LB-IF\\{\scriptsize$n=100$}} & Family & Example \\
\midrule
\multicolumn{8}{@{}l}{\textit{No task labels supplied at inference}} \\
Generic seed & --- & 26.0 & 32.0 & 76.7 & 75.8 & 52.6 & 51.1 \\
MoP & $\approx$1k & 31.2 & 16.0 & 78.5 & 79.2 & 51.2 & 54.0 \\
GRPO & 18k & 28.5 & 33.3 & 76.3 & 78.1 & 54.0 & 52.5 \\
GRPO ($5\times$ budget) & 90k & 26.2 & 39.3 & 78.9 & 80.6 & 56.3 & 53.0 \\
\fpa-ALL & 18k & 54.1 & 36.7 & 67.8 & \textbf{91.4} & 62.5 & 63.6 \\
\rowcolor{blue!7}
\agepa & 18k & \textbf{58.2} & \textbf{47.3} & \textbf{91.8} & 85.0 & \textbf{70.6} & \textbf{73.2} \\
\midrule
\multicolumn{8}{@{}l}{\textit{Reference: task labels supplied at inference}} \\
Per-family \fpa & $4{\times}4.5$k & 63.5 & 42.0 & 92.4 & 94.2 & 73.0 & 77.0 \\
\bottomrule
\end{tabular}
\end{table}

\paragraph{The routed library achieves a higher overall score.}
\agepa reaches $70.6$, versus $62.5$ for \fpa-ALL and $73.0$ for the label-routed reference (Table~\ref{tab:main_results}).
All use the same nominal aggregate budget. The advantage over \fpa-ALL also holds under example weighting.

\paragraph{The mean advantage is not confined to PUPA.}
PUPA contributes $6.0$ of the $8.1$-point Qwen advantage.
Reaveraging the other three tasks for the same selected candidates gives $63.5$ versus $60.7$; the corresponding GPT-4.1 means are $65.0$ versus $62.1$.
The difference stays positive for every excluded family, without candidate reselection (Appendix~\ref{app:score-sensitivity}).

\paragraph{The advantage emerges early in search.}
At 5,018 scored calls, \agepa reaches a validation mean of $67.5$, above the best of \fpa-ALL ($65.6$) and the 18,000-call GRPO run ($60.0$).
It finishes at $74.0$ (Figure~\ref{fig:main-story}a).
These curves count scored calls, not total compute.

\paragraph{Longer GRPO training leaves a substantial gap.}
GRPO reaches $56.3$ at the retrospective 90k-call cutoff ($5\times$ budget), up from $54.0$ at 18k. Validation selects both checkpoints (Appendix~\ref{app:grpo-budget}).

\paragraph{The higher mean includes a different balance across capabilities.}
\fpa-ALL leads on LiveBench-IF but scores $67.8$ on PUPA, below the common seed reference ($76.7$). \agepa improves all four task families over the common seed reference.

\subsection{Results with GPT-4.1}
\label{sec:results:actor-transfer}

\paragraph{The routed library remains competitive with a stronger actor.}
With GPT-4.1 execution and GPT-5.5 reflection, \agepa scores $71.5$, versus $69.8$ for \fpa-ALL and $74.9$ for the label-routed reference (Table~\ref{tab:gpt41}). Its four specialists match all 651 post-hoc task labels.

\begin{table}[H]
\centering
\caption{\textbf{GPT-4.1 narrows the advantage over single-program evolution.} Conventions follow Table~\ref{tab:main_results}. Resumptions and a judge outage limit the evidence (Appendix~\ref{app:validity}).}
\label{tab:gpt41}
\small
\setlength{\tabcolsep}{8.2pt}
\begin{tabular}{@{}>{\columncolor{white}[0pt][\tabcolsep]}lrrrrrr>{\columncolor{white}[\tabcolsep][0pt]}r@{}}
\toprule
 & Budget & \multicolumn{4}{c}{Test score ($\times100$) $\uparrow$} & \multicolumn{2}{c}{Mean ($\times100$) $\uparrow$} \\
\cmidrule(lr){3-6}\cmidrule(l){7-8}
Method & (calls) & \shortstack{HotpotQA\\{\scriptsize$n=300$}} & \shortstack{AIME\\{\scriptsize$n=30$}} & \shortstack{PUPA\\{\scriptsize$n=221$}} & \shortstack{LB-IF\\{\scriptsize$n=100$}} & Family & Example \\
\midrule
\multicolumn{8}{@{}l}{\textit{No task labels supplied at inference}} \\
Generic seed & --- & 53.3 & 30.7 & 80.5 & 75.3 & 59.9 & 64.9 \\
MoP & $\approx$1k & 47.3 & 27.3 & 75.4 & 73.3 & 55.8 & 59.9 \\
\fpa-ALL & 18k & 58.1 & 36.0 & \textbf{93.0} & 92.3 & 69.8 & 74.2 \\
\rowcolor{blue!7}
\agepa & 18k & \textbf{62.5} & \textbf{40.0} & 90.8 & \textbf{92.6} & \textbf{71.5} & \textbf{75.7} \\
\midrule
\multicolumn{8}{@{}l}{\textit{Reference: task labels supplied at inference}} \\
Per-family \fpa & $4{\times}4.5$k & 71.7 & 39.3 & 92.6 & 96.1 & 74.9 & 81.0 \\
\bottomrule
\end{tabular}
\end{table}

\begin{figure}[H]
\centering
\includegraphics[width=\linewidth]{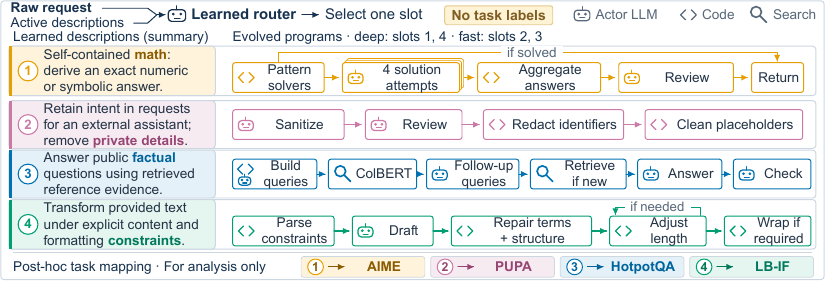}
\caption{\textbf{The final harness learns different procedures.} Qwen3-8B candidate 74 pairs descriptions (left) and programs (right), with post-hoc task labels below. Slot 5 is inactive; fallback is fixed.}
\label{fig:final-library}
\end{figure}

\subsection{The learned division changes how requests are executed}
\label{sec:results:library}

\begin{wraptable}[12]{r}{0.43\textwidth}
\centering
\vspace{-0.5cm}
\caption{\textbf{Routing matches the task partition.} Family agreement (\%) under a post-hoc majority-family mapping.}
\label{tab:router}
\small
\setlength{\tabcolsep}{3.5pt}
\begin{tabular}{@{}lrr@{}}
\toprule
Family & MoP & Ours \\
\midrule
HotpotQA & 94.7 & 100.0 \\
AIME & 100.0 & 100.0 \\
PUPA & 88.2 & 100.0 \\
LB-IF & 97.0 & 100.0 \\
\midrule
Overall & 93.1 & 100.0 \\
\bottomrule
\end{tabular}
\end{wraptable}

\paragraph{Description and router edits change program use.}
The final router matches the post-hoc family mapping on all 651 test requests, with no fallback or inactive-slot selection (Table~\ref{tab:router}). This measures partition agreement, not selection of the best-performing expert for each request.
A description-only edit adds retrieval examples such as ``animals and breeds,'' raising use of the unchanged program from 34/45 to 44/45 factual-QA validation requests, but also attracting two constrained rewrites.
A router-only edit then prioritizes the requested operation over topic words: retrieval covers 45/45 factual-QA requests, and both rewrites return to the fallback (Appendix~\ref{app:description-evolution}).
All specialist programs remain unchanged.
These edits change coverage and resolve overlap; they do not isolate an answer-quality gain.

\paragraph{Specialists learn different procedures.}
Figure~\ref{fig:final-library} pairs descriptions with programs: mathematics can answer directly or review four attempts; rewriting repairs drafts in code (Appendix~\ref{app:library}).

\subsection{Search combines improvements from different branches}
\label{sec:results:evolution}

\begin{wrapfigure}{r}{0.47\linewidth}
\vspace{-\intextsep}
\centering
\includegraphics[width=\linewidth]{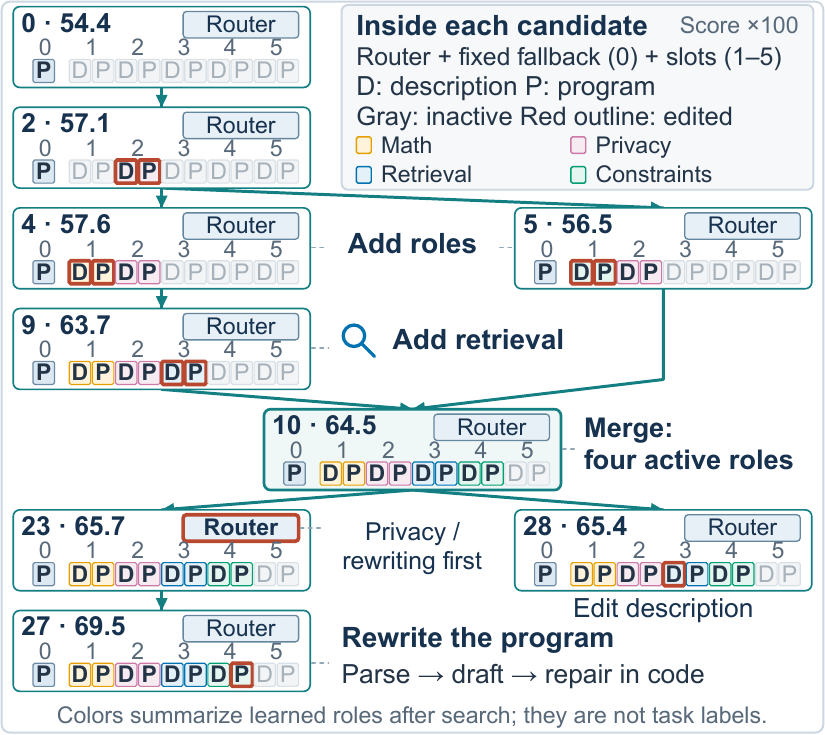}
\caption{\textbf{Evolving libraries.} Nodes show validation means; arrows link parents.}
\label{fig:early-evolution}
\end{wrapfigure}

\paragraph{Search creates roles, then refines and combines them.}
Figure~\ref{fig:early-evolution} follows activation and merging into four active roles at candidate 10, followed by router, description, and program edits.
Later, candidate 46 combines mathematics from 37 with routing, privacy, and rewriting from 31, reaching a validation mean of $73.05$.
Three pairs persist in the final library (full graph in Appendix~\ref{sec:appendix:tree}).

\begin{wrapfigure}[15]{r}{0.3\textwidth}
\vspace{-\intextsep}
\centering
\includegraphics[width=\linewidth]{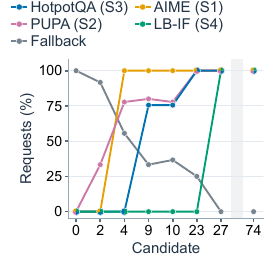}
\caption{\textbf{Slot use.} 45 requests per task; 180 for fallback. Slots are mapped to tasks after search.}
\label{fig:slot-evolution}
\end{wrapfigure}

\paragraph{Active slots need not receive requests.}
Slot 4 is active at candidate 10 but receives no executions (Figure~\ref{fig:slot-evolution}). Fallback use falls from $36.7\%$ to $25\%$ after a router edit ($10\to23$), then to zero after a slot-4 program edit ($23\to27$). Each specialist then handles $100\%$ of its task's requests. Lines trace one ancestry path; missing records separate selected candidate 74. These rates measure allocation, not answer accuracy.

\WFclear

\paragraph{Request overlap addresses a concrete ambiguity in merging.}
Specialists can occupy different slots across branches. In merge $55+59\to64$, request overlap aligns four roles; shared-request scores select two pairs from each parent. The child passes the five-request check and scores $72.0$ on full validation, versus $68.4$ and $68.3$ for its parents. A proposal from 55 and 61 fails its check. Accepted merges can still regress on individual families (Appendix~\ref{app:main-merge}).

Across all 26 accepted merges, ten retain active description--program pairs unique to each parent.
All ten have higher full-validation means than both parents, with a median gain of $3.2$ points; two are ancestors of the selected candidate.
Thirteen archived children match a parent, and three differ only in inactive code (Appendix~\ref{app:main-merge}).
These are selected observations, not controlled estimates of the merge operator's effect.

\WFclear
\subsection{How search distributes local edits}
\label{sec:results:edit-allocation}

\begin{wraptable}[10]{r}{0.48\textwidth}
\vspace{-\intextsep}
\centering
\caption{\textbf{Local edit counts.} All branches of the Qwen3-8B run; merges excluded.}
\label{tab:local-edits}
\small
\setlength{\tabcolsep}{4pt}
\begin{tabular}{@{}lrr@{}}
\toprule
Local edit & Proposed & Accepted \\
\midrule
Activate specialist & 22 & 16 \\
Rewrite program & 38 & 27 \\
Rewrite router & 12 & 8 \\
Rewrite description & 4 & 1 \\
\midrule
Total & 76 & 52 \\
\bottomrule
\end{tabular}
\end{wraptable}

\paragraph{Most local proposals create or refine programs.}
In the main Qwen3-8B run, 60 of 76 local proposals ($78.9\%$) activate a specialist or rewrite a program (Table~\ref{tab:local-edits}).
The other 16 edit routing or descriptions.
Search therefore changes executable behavior as well as allocation; these counts span all branches, not a single library.

Accepted edits pass the minibatch check; none deactivate a slot.
The only accepted description edit changes coverage (Section~\ref{sec:results:library}).
Counts do not isolate performance contributions.

\paragraph{Program edits can restore execution.}
An active slot needs executable code to serve requests.
In candidate 23, slot 4's truncated source could not compile, and all 45 LiveBench-IF validation requests used the fallback.
Rewriting only this program ($23\to27$), with the router and descriptions unchanged, moved all 45 requests to slot 4 (Figure~\ref{fig:slot-evolution}); the LB-IF score rose from $61.9$ to $76.4$.
The rewrite also changed the procedure, so the gain cannot be attributed to syntax repair alone.

\section{Related Work}
\label{sec:related}

\paragraph{Reflective improvement.}
Reflexion stores verbal feedback in episodic memory \citep{reflexion2023}, while Self-Refine iteratively revises outputs \citep{selfrefine2023}.
MIPRO optimizes instructions and demonstrations across modules \citep{miprov22024}; TextGrad propagates textual feedback through computation graphs \citep{textgrad2024}.
\gepa edits prompts from execution traces and combines candidates successful on different examples \citep{gepa2025}.
We retain its reflection and per-request selection to search routed program libraries.

\paragraph{Program and tool optimization.}
ADAS searches agent code \citep{adas2024}, AFlow searches workflows \citep{aflow2025}, and AutoPDL selects prompting patterns and demonstrations in executable programs \citep{autopdl2025}.
JTPRO jointly edits global instructions and tool schemas, merging edits with the best context while preserving tool identity \citep{jtpro2026}.
Our specialists have editable implementations and evolving responsibilities; their correspondence across branches must be established before their programs are inherited.

\paragraph{Heterogeneous requests and harnesses.}
Mixture-of-Prompts routes among instruction clusters \citep{mop2024}; Causal Prompt Optimization searches query-specific prompts with an offline reward model \citep{cpo2026}.
Adaptive Auto-Harness constructs harness branches containing code and routes requests among them on open-ended streams \citep{adaptiveharness2026}.
We study a fixed mixture and jointly edit the router and specialists within each candidate, then align executed-request footprints to inherit description--program pairs across candidates.
This component correspondence is our focus within the broader setting of joint optimization and heterogeneous routing.
Appendix~\ref{app:related-comparison} compares these search spaces and related diagnostic methods.

\section{Limitations}
\label{sec:limitations}

The final method is evaluated with one seed on four fixed task families; Appendix~\ref{app:historical-seeds} reports two seeds for an earlier version. We study specialization within this mixture, not online task discovery, overlapping capabilities, or distribution shift. Some programs contain benchmark-specific branches. The comparison changes both search representation and profile allocation, so it does not isolate individual operations. Reusing validation requests for archive search and final selection can favor noisy high scores; we have not measured this effect on a fresh selection set or with repeated evaluations of the final candidates chosen by validation (Appendix~\ref{app:validation-reuse}).

Researchers saw periodic test results during development, although candidate selection used validation scores alone. PUPA can reward empty delegated requests even when runtime failures score zero; its score alone does not establish useful private delegation. Resumptions and an evaluator outage make the GPT-4.1 run supporting evidence rather than a clean replication. Appendix~\ref{app:validity} details these issues. We make no claim about latency or total compute savings.

\section{Conclusion}
\label{sec:conclusion}

\agepa makes the division of work part of program optimization. It jointly evolves a router, specialist descriptions, and executable programs, and uses observed request allocation to combine specialists across search branches. On the reported mixed workload, the resulting library outperforms the evaluated single-program baseline and approaches independently optimized, label-routed programs. The library gives one service distinct procedures for different requests.

\clearpage
\subsection*{AI use statement}
During revision, we used generative AI to clarify the method and its mathematical formulation, review the evaluation protocol, analyze and interpret existing results, and assist with writing, literature checks, figures, tables, and analysis code.
The authors take responsibility for the final paper.

\subsection*{Ethics statement}
This work evaluates generated programs on established benchmarks, including PUPA's privacy-sensitive requests.
PUPA utility alone does not establish safe or useful delegation: its metric can reward empty requests (Section~\ref{sec:limitations}).
Deployment would require separate evaluation of privacy leakage, task usefulness, and the security of generated code and tool access.
Development-time test monitoring and other evaluation limitations are documented in Appendix~\ref{app:validity}.

\subsection*{Reproducibility statement}
Appendices~\ref{app:method-details}--\ref{app:baselines} describe the search procedure, information boundaries, data splits, evaluation rules, and baseline settings. Appendices~\ref{app:merge-replay} and~\ref{app:validity} distinguish measured results from historical projections and document run limitations. The figures are generated from the recorded experiment data.

\bibliography{coral_refs}
\bibliographystyle{iclr2027_conference}

\appendix
\section{Search and merge details}
\label{app:method-details}

Search details supplement Section~\ref{sec:method}; Appendix~\ref{app:protocol} covers data, model profiles, tools, and budget.

\subsection{Candidate search}

\paragraph{Selection and reflection.}
The search starts by evaluating the seed, whose specialist slots are all inactive, on the validation set.
It stores the individual validation scores of each candidate.
Following \gepa, parent selection first removes redundant candidates from the per-example best-candidate sets, then samples candidates in proportion to the number of such sets in which they remain.
This allows different candidates to contribute strengths on different requests.
The reflection minibatch is drawn from the feedback set, independently of the validation examples used for selection.
If the parent already attains the configured perfect score on every minibatch example, the implementation skips reflection for that minibatch.
Otherwise, the action selector receives the request, routing decision, program outputs and execution trace, score, and evaluator feedback for each example.
Recent action history and slot utilization provide additional context.

\paragraph{Read access and write access.}
Table~\ref{tab:visibility} specifies what each operation can read.
These inputs are broader than the set of fields it can change.
After activation, a program rewrite returns only the target program, and a description rewrite returns only the target description.
A routing rewrite changes only the routing prompt.
Deactivation, called \texttt{quarantine} in the implementation, replaces the description with \texttt{INACTIVE} and retains the program text.
The target source remains visible during a description rewrite so the new description can reflect the existing implementation.
The prompt asks for faithful, concise descriptions with broad, distinct responsibilities rather than instance-specific roles; it does not guarantee these properties.

\paragraph{Two acceptance checks.}
A reflected child is evaluated on the same feedback minibatch as its parent and is rejected unless its score sum is strictly larger.
A merge uses five validation examples chosen from the parents' shared evaluated rows, drawing from cases favoring each parent and cases on which they tie.
Its score sum must be at least the larger of the two archived parent sums on these examples.
Passing either check triggers a full validation evaluation and insertion into the candidate pool.
There is no further requirement that the child's mean validation score exceed its parents' means.
Future parent selection depends on its individual validation scores, and the final return uses the largest mean validation score.
The core procedure is given in Algorithm~\ref{alg:agepa} in the main text.

\subsection{Specialist alignment and inheritance}

\paragraph{Footprint records.}
The merge uses the specialist that actually executed, rather than the router's raw selection.
For example, a selection that resolves to the fallback because a specialist cannot compile does not contribute evidence to that specialist's footprint.
Footprints and per-example scores are indexed by candidate content and validation-example identity.
If execution records cannot be matched to all evaluated rows, the footprint is omitted; matching row order is never assumed.

\paragraph{Initial alignment and capacity.}
The parent with the higher aggregate validation score supplies the base slot layout.
For every active specialist in each parent, the implementation finds the highest Jaccard overlap in the other parent, breaking ties in favor of the lower slot index.
It initially matches only mutual best choices with positive overlap.
Unmatched specialists from the other parent are processed in decreasing footprint size, with lower indices first on ties.
Each uses its original index if that slot is free in the base layout, or otherwise the lowest free index.
If no free slot remains, it can compete with its strongest positive-overlap base specialist that has not already been matched.
This capacity fallback need not be a mutual-best match.
A specialist without an available positive-overlap partner stays in its original parent but is omitted from the merge; slot capacity stays fixed.

\paragraph{What alignment measures.}
Footprint overlap asks whether two specialists served the same observed requests, providing an operational basis for reusing their programs.
It does not establish semantic equivalence: different routing boundaries, compilation failures, or fallback resolution can change a footprint without changing a specialist's intended role.
Mutual-best matching requires reciprocal preference for the initial pairs and can leave specialists unmatched; the capacity fallback then handles the remaining slots.
This rule does not maximize total Jaccard weight over a bipartite assignment.
The recorded examples and merge census do not establish superiority over maximum-weight matching, descriptor similarity, or hybrid rules in a controlled full search.

\paragraph{Atomic inheritance.}
The router is one merge component, and each description--program pair is another.
\gepa selects two eligible branches with a shared ancestor; neither selected parent can be an ancestor of the other.
The shared ancestor must not have a higher aggregate score than either parent.
After aligning the weaker parent to the base layout, the ancestor is left unchanged.
The usual merge eligibility check still requires at least one component that differs between the parents and matches the ancestor in one parent.
If only one parent changed a pair relative to the ancestor, its pair is inherited.
If both changed it, the specialist-specific score comparison resolves the conflict when the required evidence exists.
The description and program are copied together in every case.

\paragraph{Conflict scores and missing evidence.}
When both aligned specialists have traffic and share validation requests, their mean scores on those shared requests determine the winner; ties favor the base parent.
If only one has traffic, its pair is retained.
If score evidence is unavailable or both have traffic but their footprints are disjoint, the implementation falls back to \gepa's aggregate-parent rule, with random tie breaking.
If footprint evidence is unavailable before alignment, the original index layout is retained.
The router remains under the base layout when live specialists are reindexed or dropped, avoiding stale numeric references from the other parent's routing prompt.
If imported live specialists keep their indices, ordinary merging may also select the other router prompt.

\paragraph{Cleanup after a merge.}
After full validation, the implementation deactivates active specialists that the merged router never selected, retaining their program text.
This cleanup uses recorded routing selections, whereas the alignment footprints above use recorded executions.
It makes no additional model calls.
The archived scores and traces are those measured before cleanup; the cleaned candidate is not immediately evaluated again.
This is an approximation: no evaluated request selected a removed specialist, but changing the description list can affect the router on a later evaluation.
The reported test evaluation executes the final selected candidate, including this cleanup.

\section{Detailed optimization protocol}
\label{app:protocol}

\paragraph{Data and sampler.}
The feedback and Pareto sets each contain 45 examples from every family.
Each reflection minibatch has 24 examples, six per family.
Within LiveBench-IF, the six examples are further balanced across its configured subtasks.
The sampler uses \texttt{family} and \texttt{task} for batching, then removes both fields from router, program, and reflection inputs.

\paragraph{Evaluator feedback.}
HotpotQA feedback identifies supporting-document titles without revealing the answer.
AIME feedback includes the reference answer and a worked solution.
LiveBench-IF feedback reports satisfied and violated constraints, while PUPA returns quality and leakage judgments.
The evaluator can therefore reveal the kind of error through its feedback, even though explicit family identifiers and target specialist assignments are withheld.

\paragraph{Seed and inactive slots.}
The seed router selects among active descriptors and otherwise returns the frozen fallback \pnull.
Every nonzero descriptor initially equals \texttt{INACTIVE}.
Activation writes both a descriptor and a program because either component alone is not executable.
The fallback is a generic single-predictor DSPy program using the deep profile.

\paragraph{Profiles.}
The main actor is Qwen3-8B, and the reflection endpoint is \texttt{gpt-5.5-2026-04-24} with temperature 1.0.
The second-actor experiment uses \texttt{gpt-4.1-2025-04-14} with temperature 0.3.
PUPA's untrusted model uses the GPT-5.5 endpoint; its quality and privacy judges use local Qwen3-8B in both actor settings.
The deep executor enables Qwen3 thinking, uses temperature 0.6, and allows 8,192 output tokens.
The fast executor disables thinking, uses temperature 0.3, and allows 4,096 output tokens.
The router is always fast.
A generated program chooses its profile through a module-level \texttt{PROFILE} declaration validated by the sandbox.

\paragraph{Sandbox and tools.}
The sandbox parses and checks source before compilation, denying file and process access, arbitrary imports, dynamic execution, and unrestricted networking.
It exposes DSPy and optional \texttt{colbert\_search} over a ColBERTv2 index of Wikipedia abstracts.

\paragraph{Merge records.}
The main run records executed expert indices and scores for fully evaluated candidates. Alignment uses these execution footprints; cleanup uses routing selections. Appendix~\ref{app:method-details} details alignment, inheritance, cleanup, and fallback rules for missing evidence or limited capacity.

\paragraph{Failed execution.}
Invalid or uncompilable generated source is rejected or resolved to the frozen fallback.
Runtime exceptions from compiled specialists remain in the trace; these rollouts receive zero before family-specific evaluator dispatch, without calling the PUPA judge.

\section{Dataset details}
\label{app:data-details}

Within each family, the training, evaluation, and test examples are mutually disjoint, and no test example is ever drawn from the optimization pools.
For \textbf{HotpotQA}, the 45 training questions are sampled from the official training split, while the 45 evaluation questions and 300 test questions are disjoint samples from the official dev split.
For \textbf{AIME}, the training and evaluation examples are drawn from the 2022--2024 competitions (the pooled problems are shuffled once and halved), while the 30 problems of AIME~2025 constitute the test set; each test problem is attempted five times independently with caching disabled, and the five pass@1 scores are averaged before aggregation.
For \textbf{PUPA}, we preserve the original positional partition of its 443-example pool into 111 training, 111 evaluation, and 221 test examples, use the first 45 of each optimization slice, and retain all 221 test examples.
For \textbf{LiveBench-IF}, the 45 validation and 100 test prompts are sampled from the archived November 25, 2024 release, balanced across its four subtasks.
The 45 feedback prompts come from the rotated-out June 24, 2024 release of the same subtasks, with no source-article overlap with the validation/test pool.
The splits use the frozen parquet snapshot rather than the changing upstream release; reflection minibatches contain only feedback-set examples.
Because the evaluation set is balanced at 45 examples per family, its per-sample mean equals its family mean; Pareto candidate selection operates on the unlabeled per-example scores.

\section{Baseline and transfer configurations}
\label{app:baselines}

\paragraph{Generic seed (\pnull).}
All requests use the frozen single-predictor program from which \agepa begins.
For a common reported starting point, we use the test re-evaluation exported with the protocol-aligned GRPO run rather than mixing independent evaluations of the same seed.

\paragraph{\fpa-ALL.}
A single full program is optimized on the same mixed dataset with the same seed, nominal 18,000-call budget, feedback, and tool access.
The program can synthesize internal dispatch and control flow, but it must choose one global execution profile; the selected program uses deep.
This comparison evaluates the two implemented search systems under the same nominal scored-call cap; it does not isolate search organization from execution-profile allocation.

\paragraph{Per-family \fpa with oracle routing.}
Four full programs are optimized separately and selected at inference time using ground-truth family identity.
The programs receive the same tools as \agepa; each family receives a nominal budget of 4,500 scored calls, for an aggregate nominal budget of 18,000.
This comparison removes cross-family interference and routing error, providing a label-supervised reference for decomposition under the same aggregate search budget.
It is not a strict upper bound: independent and joint searches may discover different programs.

\paragraph{Mixture-of-Prompts (MoP).}
We reproduce the official clustering--routing--prompt-assignment pipeline of \citet{mop2024}: automatic $K$-means selection with $K\leq5$, nearest-centroid routing, APE instruction induction, and RBJS assignment.
The mixed workload uses a local Qwen3 embedding model in place of the original hosted encoder and the same task and proposal models as our other runs.
PUPA and LiveBench-IF participate in clustering and routing but provide no gold output demonstrations, because their evaluators do not expose a single reference string.
We retain MoP's native search schedule, which consumes 1,010 metric calls ($5.6\%$ of the \agepa/\fpa cap); its result is therefore an algorithmic baseline, not a compute-matched one.
The GPT-4.1 twin reuses the embedding cache and data seed, fixing the same clustering and router as Qwen (606/651 routing accuracy); only instructions and demonstrations are re-selected.

\paragraph{Group relative policy optimization (GRPO).}
We full-weight fine-tune Qwen3-8B with GRPO \citep{grpo2024} using verl on the same seed-43 feedback and Pareto splits.
The exported policy prompts use the exact DSPy ChatAdapter rendering of the generic seed \pnull, and the reward extracts the same answer-field contract with a lenient post-thinking fallback when the marker is absent.
For the 18,000-call run, 45 updates each sample 45 feedback prompts eight times (16,200 training rollouts), while validation before training and every five steps contributes ten passes over the 180-example Pareto set (1,800 calls).
Gradients use only the feedback split, and the Pareto-best checkpoint is selected for test evaluation.
A separate longer run keeps the same settings and completes 235 updates (93,240 calls). We retrospectively apply a 90,000-call cutoff and select using only validation records within it (Appendix~\ref{app:grpo-budget}).
GRPO uses the deep Qwen3-8B profile and the same family evaluators, but is a pure-generation policy without ColBERT access and uses no reflection LM; the 18,000-call comparison matches scored-rollout opportunity, not GPU-hours or auxiliary-model calls.
One residual interface difference remains: DSPy's runtime can issue a second JSONAdapter generation after a parse failure, whereas each GRPO rollout contains one generation; despite this, the aligned GRPO base matches \pnull on the test aggregate ($51.1$ micro).

\paragraph{Actor-transfer setting.}
To test whether the result depends on the Qwen executor, we repeat \agepa, \fpa-ALL, and the per-family reference with GPT-4.1 as the program actor while holding the GPT-5.5 reflection model fixed.
The deep and fast profiles use the same actor endpoint with 8,192- and 4,096-token limits, respectively; PUPA's judges remain pinned to local Qwen3-8B so that its metric is unchanged.
Because a bare non-reasoning GPT predictor follows the answer-only contract too literally on AIME, all GPT-4.1 methods start from the same generic \texttt{ChainOfThought} variant of \pnull, with the same signature and only the DSPy predictor class changed.
The data splits, feedback, candidate cap, and approximately 18,000-call aggregate budget otherwise match the main setting.
We compare methods within each actor, not across actors.

\section{Reflection information boundaries}
\label{app:actions}

Table~\ref{tab:visibility} records the information available to each reflection stage.
The action selector receives no program source.
When a source is required for an edit, only the target slot's source is exposed.

\begin{table}[ht]
\caption{\textbf{Information visible to each reflection operation.}}
\label{tab:visibility}
\centering
\small
\begin{tabular}{lccccc}
\toprule
Operation & Traces & Router & All descriptors & Target source & Sibling sources \\
\midrule
Action selection & \cmark & \cmark & \cmark & \xmark & \xmark \\
Activate slot & \cmark & \cmark & \cmark & \xmark & \xmark \\
Rewrite program & \cmark & \cmark & \cmark & \cmark & \xmark \\
Rewrite router & \cmark & \cmark & \cmark & \xmark & \xmark \\
Rewrite descriptor & \cmark & \cmark & \cmark & \cmark & \xmark \\
Quarantine slot & \xmark & \xmark & \xmark & \xmark & \xmark \\
\bottomrule
\end{tabular}
\end{table}

\section{Final candidate summary}
\label{app:library}

The archived final run records the seed-43 optimization.
The validation-selection summary reports candidate 74 with macro score 74.03.
Its four active programs occupy slots 1--4, leaving slot 5 inactive.
The router distinguishes mathematics, privacy-sensitive delegation, factual retrieval, and constrained transformation by the primary capability requested.
It prioritizes privacy-preserving external-assistant reformulation, constrained transformation of supplied text, retrieval-grounded public factual QA, self-contained exact quantitative reasoning, then the generic fallback.

The main table uses candidate 74's separate final evaluation, not the monitor's last candidate.

\subsection{Comparison with the selected single program}
\label{app:program-comparison}

The selected \fpa-ALL program also implements dispatch, multi-stage retrieval, mathematical shortcuts, and constraint repair.
Table~\ref{tab:program-comparison} compares its source with the selected library.
The distinction is how search represents and edits these choices, together with the ability to choose a profile per specialist.
The source comparison describes implemented paths, not their empirical call frequencies or which difference causes the score gap.

\begin{table}[!htbp]
\centering
\caption{\textbf{Both searches learn multi-stage procedures.} Static comparison of the validation-selected Qwen programs: \fpa-ALL candidate 21 and \agepa candidate 74. Conditional paths are condensed; shared evaluation-side models are omitted.}
\label{tab:program-comparison}
\small
\setlength{\tabcolsep}{4pt}
\renewcommand{\arraystretch}{1.1}
\newcommand{\programcell}[2]{\parbox[t]{#1\linewidth}{\raggedright\strut#2\strut}}
\begin{tabular}{@{}lll@{}}
\toprule
\programcell{0.14}{Component} & \programcell{0.40}{\fpa-ALL} & \programcell{0.40}{\agepa} \\
\midrule
\programcell{0.14}{Dispatch} & \programcell{0.40}{Python branches for privacy, writing, mathematics, retrieval; generic fallback} & \programcell{0.40}{LM router reads descriptions and selects an active program or fixed fallback} \\
\addlinespace[3pt]
\programcell{0.14}{Retrieval} & \programcell{0.40}{Heuristic and LM queries; ColBERT; code-generated bridge queries; grounded answer} & \programcell{0.40}{Heuristic and LM queries; ColBERT; evidence-conditioned follow-up; draft and LM evidence check} \\
\addlinespace[3pt]
\programcell{0.14}{Mathematics} & \programcell{0.40}{Deterministic solvers; otherwise one reasoning attempt and answer normalization} & \programcell{0.40}{Deterministic solvers; otherwise four attempts, answer aggregation, and LM review} \\
\addlinespace[3pt]
\programcell{0.14}{Privacy} & \programcell{0.40}{Regex replacement of email, phone, and a specific disability phrase} & \programcell{0.40}{LM sanitization and review; identifier cleanup; nonempty fallback request} \\
\addlinespace[3pt]
\programcell{0.14}{Writing} & \programcell{0.40}{LM draft; deterministic keyword, length, structure, and wrapper repair} & \programcell{0.40}{Constraint checklist; LM draft; deterministic repairs and word-limit loop; fixed-choice bypass} \\
\addlinespace[3pt]
\programcell{0.14}{Profile} & \programcell{0.40}{One global deep profile} & \programcell{0.40}{Deep for mathematics and writing; fast for retrieval, privacy, and routing} \\
\bottomrule
\end{tabular}
\end{table}

For example, the library's mathematics path can spend additional model calls on attempts and review, whereas the single program uses one attempt after its deterministic shortcuts.
Both writing paths repair drafts in code, and \fpa-ALL obtains the higher LB-IF score.
The privacy paths differ substantially, but archived test means cannot reveal their quality/leakage decomposition or empty-request rates.
This comparison concerns complete learned systems; it does not separate the effects of search representation, profile allocation, or individual generated procedures.

\noindent\begin{minipage}{\linewidth}
\subsection{Description and router evolution}
\label{app:description-evolution}

\noindent\begin{minipage}[t]{0.46\linewidth}
\vspace{0pt}
\paragraph{Descriptions expand coverage; router priorities separate overlapping requests.}
Figure~\ref{fig:description-evolution} follows the recorded branch \mbox{$10\to28\to36$}.
First, only the retrieval description changes, adding explicit examples such as ``animals and breeds'' to its public-factual scope.
Next, only the router prompt changes: a supplied article with rewriting constraints belongs to the rewriting specialist, even when it mentions public facts.
All five specialist programs remain identical.

After the router edit, the two rewriting requests return to the fallback.
These execution counts use task labels only for analysis.
They describe one trajectory, not initial router choices or isolated answer-quality effects: the newly retrieved factual request scores lower, and both rewriting scores are unchanged.
\end{minipage}\hfill
\begin{minipage}[t]{0.51\linewidth}
\vspace{0pt}
\centering
\includegraphics[width=\linewidth]{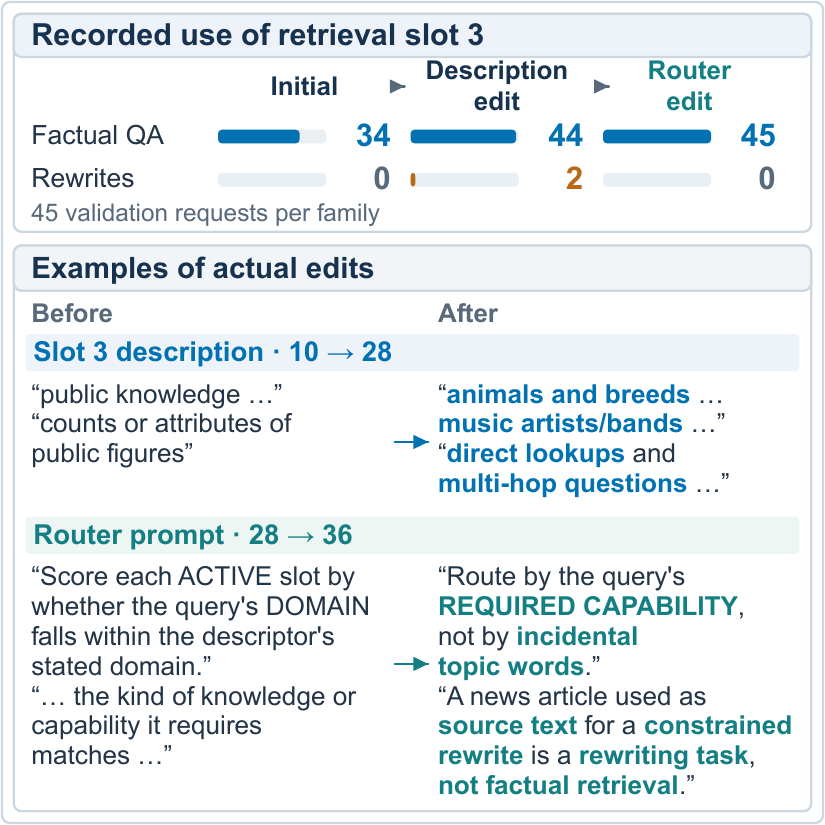}
\figurecaption{\textbf{Edits change program use.} Top: recorded executions. Bottom: before/after description and router excerpts; ellipses mark omissions.}
\label{fig:description-evolution}
\end{minipage}
\end{minipage}
\par\medskip

\section{Additional test results}
\label{app:additional-results}

Tables~\ref{tab:main_results} and~\ref{tab:gpt41} report family and per-example means rounded after aggregation. The per-example mean weights HotpotQA, AIME, PUPA, and LiveBench-IF by 300, 30, 221, and 100. AIME's five attempts are averaged per problem and do not increase its weight.

\paragraph{Seed evaluations.}
The common Qwen3-8B seed row uses the evaluation exported with the aligned GRPO run. Its family mean is $52.6081$, and the selected GRPO checkpoint's mean is $54.0367$.
The independent seed evaluation in the \agepa archive has a mean of $52.1220$, reflecting sampling variation in evaluating the same program.
The common GPT-4.1 seed row is taken from the \agepa archive. MoP's independent GPT-4.1 seed evaluation yields $57.6$ family mean and $62.6$ per-example mean; its selected prompts yield $55.8$ and $59.9$. A shared displayed seed is a reporting reference, not evidence that all independent initial evaluations were numerically identical.

\subsection{Sensitivity to individual task families}
\label{app:score-sensitivity}

The reported advantage is not produced solely by one task's inclusion in the mean.
For each actor, Table~\ref{tab:score-sensitivity} reaverages the archived task scores of the same validation-selected candidates after excluding one family at a time.
No candidate is reselected or reevaluated, and every search still used all four families.
PUPA contributes $6.00$ of the $8.11$-point Qwen difference, or $74.0\%$.
Excluding PUPA leaves $2.81$ points for Qwen and $2.91$ for GPT-4.1, whose PUPA score is below the baseline.
These are reporting checks on fixed runs, not searches trained with fewer task families.

\begin{table}[!htbp]
\centering
\caption{\textbf{The mean advantage remains after excluding any one family.} Scores ($\times100$) use fixed final candidates and equally weight the remaining families. $\Delta$ is Ours minus FPA.}
\label{tab:score-sensitivity}
\small
\setlength{\tabcolsep}{8pt}
\begin{tabular}{@{}lrrrrrr@{}}
\toprule
 & \multicolumn{3}{c}{Qwen3-8B} & \multicolumn{3}{c}{GPT-4.1} \\
\cmidrule(lr){2-4}\cmidrule(l){5-7}
Excluded family & FPA & Ours & $\Delta$ & FPA & Ours & $\Delta$ \\
\midrule
None & 62.47 & 70.58 & +8.11 & 69.84 & 71.46 & +1.62 \\
HotpotQA & 65.27 & 74.71 & +9.44 & 73.76 & 74.44 & +0.68 \\
AIME & 71.07 & 78.32 & +7.26 & 81.12 & 81.95 & +0.83 \\
PUPA & 60.69 & 63.50 & +2.81 & 62.12 & 65.03 & +2.91 \\
LiveBench-IF & 52.83 & 65.76 & +12.93 & 62.36 & 64.43 & +2.07 \\
\bottomrule
\end{tabular}
\end{table}

The final-test exports retain task means but no per-request score vectors, so they do not support paired confidence intervals or a breakdown of PUPA's quality, leakage, and empty-request rates.
The selected programs can be inspected, but source code cannot establish these empirical rates.
These reaggregations do not remove search randomness, development-time test exposure, profile differences, or the judge outage and interruptions in the GPT-4.1 run.

\newpage
\subsection{GRPO with a larger training budget}
\label{app:grpo-budget}

GRPO still trails program search at a retrospective $5\times$ budget cutoff. The longer run keeps the 45-step run's settings, targets 500 updates, and stops at 235 (93,240 scored calls). For Table~\ref{tab:main_results}, we use its recorded prefix through update 227:
\[
\underbrace{227\times45\times8}_{81{,}720\text{ training calls}}
+\underbrace{46\times180}_{8{,}280\text{ validation calls}}
=90{,}000=5\times18{,}000.
\]
Validation covers 180 requests before training and every five updates; test calls are excluded. Only the 46 validation records through step 225 enter the 90k selection; later records are excluded. Their highest mean selects step 180, as does the full run. We reuse its existing test evaluation; no training or evaluation is rerun, and no new validation pass is added at step 227.

\begin{table}[H]
\centering
\caption{\textbf{Validation and test scores favor different checkpoints.} All scores are multiplied by 100. Both the 90k cutoff and full-run validation select step 180. Steps 115 and 235 are post-hoc diagnostics; step 235 lies beyond 90k. Means follow Table~\ref{tab:main_results}.}
\label{tab:grpo-budget}
\small
\setlength{\tabcolsep}{3pt}
\begin{tabular*}{\linewidth}{@{\extracolsep{\fill}}lrrrrrrr@{}}
\toprule
 & & \multicolumn{4}{c}{Test score} & \multicolumn{2}{c}{Test mean} \\
\cmidrule(lr){3-6}\cmidrule(l){7-8}
Checkpoint & Val. mean & HotpotQA & AIME & PUPA & LB-IF & Family & Example \\
\midrule
Seed & 50.4 & 26.6 & 36.7 & 78.5 & 74.3 & 54.0 & 52.0 \\
115 & 60.2 & 28.5 & 36.0 & 80.2 & 81.9 & 56.6 & 54.6 \\
\textbf{180 (selected)} & \textbf{62.8} & 26.2 & 39.3 & 78.9 & 80.6 & 56.3 & 53.0 \\
235 (final) & 59.7 & 27.9 & 40.7 & 76.8 & 84.1 & 57.4 & 53.7 \\
\bottomrule
\end{tabular*}
\end{table}

Table~\ref{tab:grpo-budget} evaluates its seed independently of Table~\ref{tab:main_results}; checkpoints are selected without test scores.

\section{Recorded merges and component inheritance}
\label{app:main-merge}

\paragraph{Some accepted merges combine programs; others preserve a parent.}
The main run constructs 47 merge proposals; 26 pass the five-request check and receive full validation.
We compare each archived child with both parents using exact source and description equality, independently of slot index.
Ten children retain at least one active description--program pair unique to each parent, thirteen match a parent's complete serialized fields, and three differ only in stored code for inactive slots (Table~\ref{tab:recorded-merges}).
There are no children whose only difference is the router prompt.

\begin{table}[!htb]
\centering
\caption{\textbf{Ten accepted merges combine distinct active pairs from both parents.} All 26 accepted merges in the main run. $\Delta$ is the recorded full-validation mean minus the stronger parent's mean, multiplied by 100. Classes describe the archived source after cleanup; scores precede cleanup.}
\label{tab:recorded-merges}
\small
\setlength{\tabcolsep}{5pt}
\begin{tabular}{@{}lrrrrl@{}}
\toprule
Archived child & Count & Higher & Lower & Median $\Delta$ & Range of $\Delta$ \\
\midrule
Active pairs from both parents & 10 & 10 & 0 & +3.20 & $[+0.34,+6.54]$ \\
Identical to a parent & 13 & 7 & 6 & +0.08 & $[-2.05,+2.55]$ \\
Only inactive code differs & 3 & 1 & 2 & $-2.85$ & $[-3.05,+0.49]$ \\
\midrule
All accepted merges & 26 & 18 & 8 & +0.67 & $[-3.05,+6.54]$ \\
\bottomrule
\end{tabular}
\end{table}

\begin{figure}[!htb]
\centering
\includegraphics[width=\linewidth]{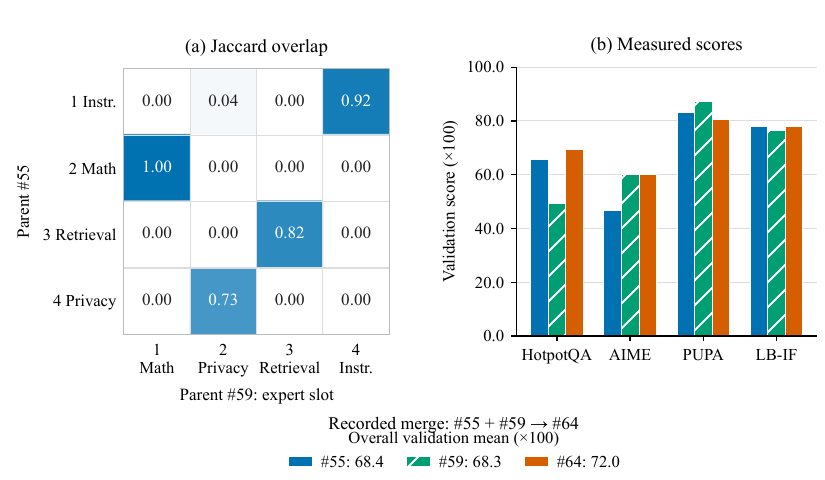}
\caption{\textbf{Observed request overlap aligns specialists across different slot layouts.} Left: Jaccard overlap of the two parents' recorded validation footprints. Right: measured family scores for both parents and their merged child. Large off-diagonal entries show that slot position alone does not identify a responsibility. Task labels only annotate the plot; Instr. means instruction following.}
\label{fig:jaccard-merge-example}
\end{figure}

\begin{figure}[!htb]
\centering
\includegraphics[width=\linewidth]{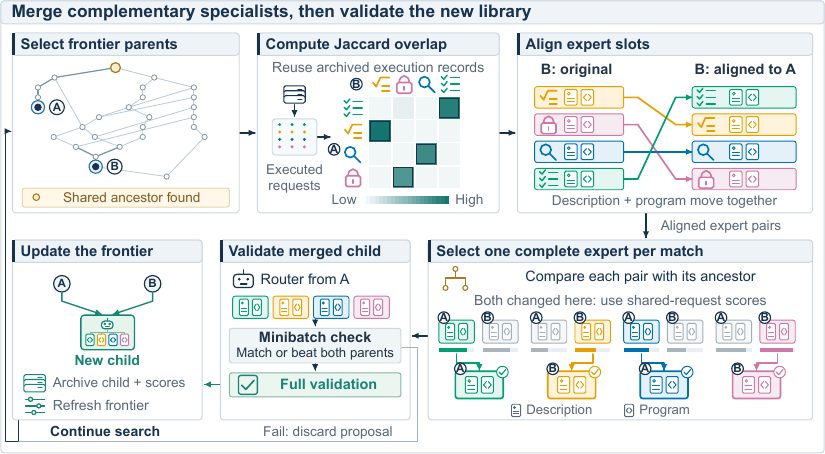}
\caption{\textbf{Merging aligns specialists, selects components, and evaluates a new library.} Archived request overlap aligns parents 55 and 59; shared-request scores select description--program pairs when both changed. Validation precedes archiving and frontier refresh. Colors show post-hoc roles.}
\label{fig:evolution-path}
\end{figure}

All ten combinations have higher recorded validation means than both parents; their median difference is $3.20$ points.
These ten events produce eight distinct serialized children; candidates 6/8 and 49/54 repeat the same respective child.
Two, candidates 10 and 46, lie on the selected candidate's ancestry.
Candidate 46 retains mathematics from 37 and routing, privacy, and writing from 31.
Its mathematics, privacy, and retrieval pairs remain unchanged in the final candidate 74; the writing program is subsequently edited.
These observations establish that search reused components from different branches; they do not show those merges were counterfactually necessary.
All comparisons concern proposals that passed the five-request check and reuse separate stochastic parent evaluations, so they do not estimate an unbiased causal effect of merging.

\paragraph{Cleanup and repeated source require separate accounting.}
The launch log records post-validation cleanup at candidates 12, 25, 31, 39, and 46, each deactivating slot 5.
Thus thirteen identical archived children do not all correspond to unchanged source at evaluation: candidate 39 becomes identical only after cleanup.
The other twelve have no cleanup change; six score above the stronger parent and six below, with a median difference of $-0.05$ points.
Some archived parent scores also precede cleanup, so this subset is not a pure estimate of sampling noise.
Candidate 74 has no recorded cleanup change and matches parent 73, despite their different measured validation means.
This distinguishes source reuse from a new program improvement without assuming evaluations of identical source return identical scores.

\paragraph{Alignment resolves a real difference in slot identity.}
The main run provides a direct example of why specialist identity should be separated from slot position. Candidate 55 places instruction following, mathematics, retrieval, and privacy in slots 1--4, respectively; candidate 59 places mathematics, privacy, retrieval, and instruction following in those slots.
Figure~\ref{fig:jaccard-merge-example} shows their recorded footprint overlap and the evaluated child, candidate 64. Matching uses executed validation requests, without task-family labels; the family names in the figure are post-hoc annotations.
The child reaches a measured validation mean of $71.99$, compared with $68.43$ and $68.29$ for its parents.
Its PUPA score is $80.74$, below both parents ($83.38$ and $87.35$): choosing components from archived scores does not guarantee that the merged library preserves every family score.
These are observations from the actual search, unlike the score projections in Appendix~\ref{app:merge-replay}. They do not by themselves estimate what a different merge operator would have achieved with the same parents.
Figure~\ref{fig:evolution-path} traces parent selection and alignment, child construction, validation, and archiving.

\section{Historical runs and merge replay}
\label{app:merge-replay}

\subsection{An earlier version with two seeds}
\label{app:historical-seeds}

An earlier version (v4) improves all three reported families over its own initial program in both archived runs (Table~\ref{tab:historical-seeds}).
These Qwen3-8B runs use seeds 42 and 43, index-based merging, and zero-traffic cleanup, before the adoption of footprint alignment and corrected failure scoring.

\begin{table}[!htb]
\centering
\caption{\textbf{Historical runs improve over their own initial programs.} Results for predecessor v4, not repeated runs of the final method. Scores are multiplied by 100; the mean excludes PUPA. $\Delta$ compares each selected program with its own initial program.}
\label{tab:historical-seeds}
\small
\setlength{\tabcolsep}{6pt}
\begin{tabular}{@{}llrrrrr@{}}
\toprule
Seed & Program & HotpotQA & AIME & LB-IF & Mean & $\Delta$ \\
\midrule
42 & Initial & 28.7 & 31.3 & 67.6 & 42.56 & --- \\
42 & Selected & 60.0 & 40.0 & 71.5 & 57.16 & +14.60 \\
\addlinespace[3pt]
43 & Initial & 25.4 & 34.0 & 70.3 & 43.24 & --- \\
43 & Selected & 62.6 & 41.3 & 80.2 & 61.38 & +18.14 \\
\bottomrule
\end{tabular}
\end{table}

PUPA is excluded because the earlier scoring rule could reward empty outputs after execution errors; a saved diagnostic confirms this problem in the seed-43 privacy specialist.
Both runs still optimized and selected candidates using all four families, so this exclusion does not remove PUPA's influence on search.
We reuse the archived final evaluations without reselection or rescoring.

Seed 42 resumed from a candidate budget of 67 to 100; seed 43 was configured for 100 from the start.
The seed changes search and HotpotQA/LB-IF sampling; AIME and PUPA splits stay fixed.
These records show improvements in two historical searches, not multi-seed validation of the final algorithm or its advantage over \fpa-ALL.

\subsection{Replay of historical merges}

\paragraph{What the replay measures.}
We inspect whether two merge strategies retain different specialists when given the same historical parents and their common ancestor.
The index strategy uses the original merge procedure; the Jaccard strategy adds both specialist alignment and conflict resolution based on specialist scores.
The replay reconstructs each slot's footprint by identifying its task family from the descriptor and assigning that family's validation examples to it.
These reconstructed footprints use task labels for post-hoc analysis; they are distinct from the actual routing footprints recorded during the main optimization run.
The resulting children are not executed.
Instead, the replay estimates a child's score by combining the archived task-family scores of the parents whose programs it retains.
A child without a specialist for every family has no complete projected mean.
These projections do not measure changes in routing or stochastic program execution.

\paragraph{Coverage and counts.}
The three historical runs contain 120 logged merge events.
The replay skips 86 events whose slot layouts cannot be mapped unambiguously to task families, and the Jaccard strategy declines 13 more after alignment.
Of the 21 remaining events, 10 produce identical children under both strategies, 10 improve the projected mean under Jaccard, and one restores the retrieval specialist lost by the index strategy (Table~\ref{tab:merge-case-study}).
Among the 20 events with complete projections under both strategies, none has a lower projected mean under Jaccard.
This statement applies only to these projections and successfully replayed events.
It does not imply that every family score improves relative to a historically measured child: in the example table, the measured index child's PUPA score is $88.1$, whereas the Jaccard projection is $86.5$.

\begin{table}[ht]
\centering
\caption{\textbf{Index-based merging can lose a retrieval specialist.} Scores are multiplied by 100. Parent and index-child scores are measured; the aligned-child row is projected from retained parent programs. Candidate indices refer to a historical run, not the main run.}
\label{tab:merge-case-study}
\small
\setlength{\tabcolsep}{4pt}
\begin{tabular}{@{}lrrrrr@{}}
\toprule
 & HotpotQA & AIME & PUPA & LB-IF & Mean \\
\midrule
Parent 52 & 62.8 & 35.6 & 86.5 & 89.9 & 68.7 \\
Parent 66 & 73.6 & 33.3 & 88.3 & 73.7 & 67.2 \\
Index child (measured) & 38.3 & 35.6 & 88.1 & 89.9 & 63.0 \\
Aligned child (projected) & 73.6 & 35.6 & 86.5 & 89.9 & 71.4 \\
\bottomrule
\end{tabular}
\end{table}

\paragraph{Relation to end-to-end results.}
The historical runs differ in random seed or failure scoring, and the final configuration changes both the merge strategy and the treatment of runtime exceptions.
Their end-to-end differences therefore cannot be attributed to the merge strategy alone.
The replay shows that slot-index matching can discard a specialist; it does not isolate the contributions of alignment or conflict resolution to final test scores.

\section{Full optimization graph}
\label{sec:appendix:tree}

\begin{figure}[H]
\centering
\includegraphics[width=\textwidth]{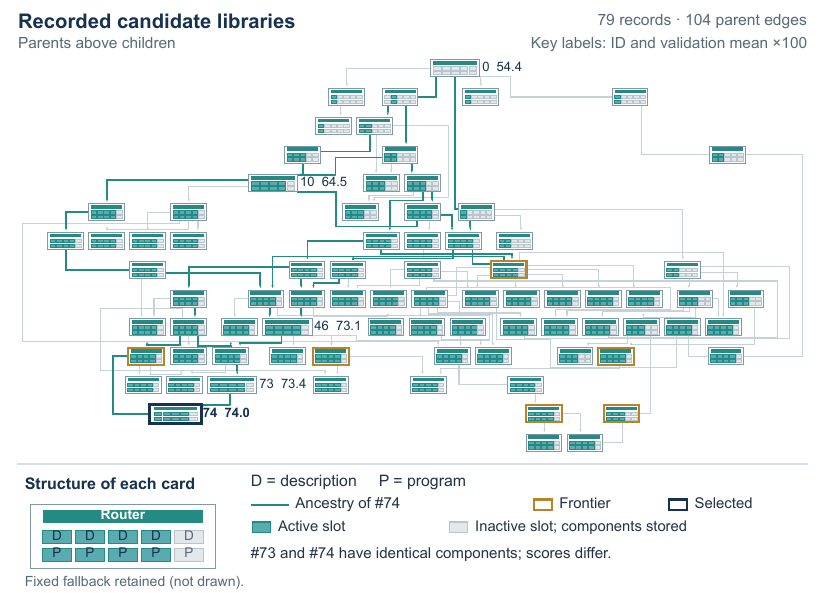}
\caption{\textbf{Search evolves entire candidate libraries.} All 79 records and 104 parent edges are shown. Each card contains a router above five description/program pairs; gray pairs are inactive but stored. The fixed fallback is retained but omitted. Teal edges trace candidate 74's ancestry; gold borders mark recorded final-frontier membership. Candidate 74 has a separate selected border; its frontier status is unrecorded. Labels give IDs and validation means. Candidates 73 and 74 contain identical components despite their different scores.}
\label{fig:evolution-tree-full}
\end{figure}

Figure~\ref{fig:evolution-tree-full} reports the complete recorded search graph; Figure~\ref{fig:early-evolution} expands an early subgraph with component-level annotations.
The graph contains 79 candidate evaluations, including the seed and 26 two-parent merges, but only 64 distinct serialized candidates.
The highlighted ancestry includes every recorded parent of the selected candidate. A parent link records a search operation; it does not imply that the child retains a changed component from that parent.

\paragraph{Validation gains need not imply a new program.}
The selected candidate 74 has the same serialized components as candidate 73, despite its higher measured validation score ($74.03$ versus $73.38$).
The merge retains the library, so selection reflects a new evaluation, not a structural change.

\section{Compute and logged usage}
\label{app:compute}

The main run uses candidate budget 100 and records 123 optimization steps, 79 candidates, and 18,103 metric calls.
The reflection counters record 152 completions and 11,913,344 tokens for \agepa, compared with 92 completions and 5,382,576 tokens for \fpa-ALL.
Both use the same provider-usage accumulator; \agepa includes action selection and target editing, while its aligned merge makes no reflection call.
The counters exclude failed attempts without reported usage, so they measure recorded completions rather than all network requests.
The task-model log reports a retained history of 10,000 calls and 10,149,552 tokens for the deep-profile model, not total actor usage.
DSPy's history is capped at 10,000 entries, and this callback does not cover the fast-profile, router, evaluation-twin, or PUPA auxiliary models.
Complete actor-token totals are therefore unavailable; these history counts cannot support a total-compute comparison.

The shorter GRPO run completes 45 update steps on six GPUs and takes approximately five hours.
Its 18,000 scored calls comprise 16,200 sampled training rollouts and 1,800 Pareto-validation generations; it makes no reflection-model calls.
The policy prompt is the DSPy $P_0$ ChatAdapter rendering, and its reward extracts the corresponding answer field with a lenient fallback.
The pre-specified rule selects validation-best step 30 for held-out evaluation.
The full longer run completes 235 updates in about 18 hours 46 minutes and consumes 93,240 scored calls. The reported 90,000-call cutoff uses its recorded prefix; validation selects step 180 under both budgets (Appendix~\ref{app:grpo-budget}).
This matches scored-rollout opportunity, not FLOPs, tokens, or wall-clock time.

The test-monitor callback evaluates every fifth accepted candidate.
For AIME, it repeats all 30 questions five times with cache disabled.
These repeated samples explain why the callback's raw aggregate weights AIME as 150 evaluations.
The tables average AIME attempts within each problem first, then report family means or per-example means as labeled.

\paragraph{GPT-4.1 actor transfer.}
The transfer runs reuse the seed-43 splits and replace the actor with GPT-4.1 while retaining GPT-5.5 for reflection and local Qwen3-8B for PUPA judging.
The program-search methods start from the same generic \texttt{ChainOfThought} seed described in Appendix~\ref{app:baselines}.
The selected \agepa candidate is index 75 with validation macro $70.04$; the selected \fpa-ALL candidate is index 15 with validation macro $70.86$.
The four per-family \fpa runs each receive a nominal budget of 4,500 scored calls and yield held-out scores of $71.66$, $39.33$, $92.59$, and $96.09$ for HotpotQA, AIME, PUPA, and LiveBench-IF, respectively.
\agepa's selected library has four active family-aligned slots, one inactive slot, and a 651/651 post-hoc routing audit.

\section{Validity and evaluation limitations}
\label{app:validity}

\paragraph{Harness metadata and fixed mixture.}
The four families and their proportions remain fixed during optimization.
The benchmark harness uses source metadata to choose evaluators and balance reflection minibatches; the candidate, router, executor, and reflection LM never receive it.
The experiment does not test unsupervised task discovery, online adaptation, or unseen families.

\paragraph{Single main seed and coupled method changes.}
The final run uses seed 43.
Alignment and corrected failure scoring were introduced together; earlier runs merged by index.
Replay diagnoses merges but cannot isolate version effects or replace a prospective multi-seed ablation.

\paragraph{Repeated validation and selection.}
\label{app:validation-reuse}
Each of the 180 validation requests was scored in 79 full-candidate evaluations for the main Qwen \agepa run and 76 for \fpa-ALL, including the seed.
\agepa also ran 47 five-request merge checks, adding 235 scores; 26 passed and received full validation.
Per-request validation counts were 79/80/86 (minimum/median/maximum), versus 76/76/76 for \fpa-ALL.
The resulting 14,455 and 13,680 validation scores, plus 3,648 and 4,416 feedback scores, account for the reported budgets of 18,103 and 18,096 calls.
Merge parents use stored scores; archive selection, alignment, and conflict resolution also reuse validation records. The logs do not count every such comparison.
These are scoring invocations, not independent LM samples: actor response caching remained enabled during optimization.
Final selection reuses the same validation set; we have no fresh selection set or repeated finalist evaluation to estimate the resulting selection optimism.

\paragraph{PUPA metric incentives.}
The implementation fix prevents an exception from being rewarded, but it does not alter the benchmark itself.
A program that deliberately emits an empty delegation request can still score well because it reveals nothing while the local response path retains the original query.
All methods face the same metric, and the final \agepa program implements explicit redaction, but this incentive limits what the PUPA score alone establishes about useful delegation.
The PUPA scorer divides the judge's reported number of leaked items by the number of annotated private items without clipping the ratio. If the judge overcounts, leakage can exceed one and the resulting utility can be negative. The merge example includes such scores: candidate 59 has one negative PUPA score, and candidate 64 has two. All reported aggregates preserve the archived values; these are judge-derived utility scores, not execution-failure penalties.

\paragraph{MoP adaptation and budget.}
Our MoP reproduction replaces its original hosted embedding model, applies it to families without gold demonstration strings, and retains its native budget rather than matching \agepa's 18,000-call cap.
This tests transfer of the published prompt-mixture recipe to an executable mixed workload, not performance at matched compute.

\paragraph{GRPO budget and interface.}
The shorter GRPO run matches the nominal scored-call budget; the longer-run result uses a retrospective $5\times$ cutoff. Neither matches wall-clock compute or auxiliary calls: GRPO full-weight trains on six GPUs without a reflection LM, whereas \agepa calls a remote reflection model and executes generated programs.
Without retrieval tools, these runs test a shared weight-space policy on heterogeneous requests at two rollout budgets; they do not establish that RL with an agentic tool interface is intrinsically weaker.

\paragraph{Test monitoring during development.}
Candidate selection within a run uses only $\dpareto$, and test scores are not consumed by the optimizer.
However, a callback evaluated every fifth accepted candidate, and those curves were visible during method development.
Consequently, the test sets were not isolated from researcher decisions during development.

\paragraph{GPT-4.1 transfer-run integrity.}
The actor-transfer run resumed twice, after a pathological regular expression and a runaway loop.
A temporary PUPA-judge outage in the second segment assigned zero PUPA vectors to candidates 15--22.
Selected candidate 75 was created after recovery, but the outage may have altered the search trajectory.
Resumption reset the merge counter, and the run exceeded the configured merge cap.
This run provides corroborative actor-transfer evidence, not a clean replication or a search-efficiency comparison.

\paragraph{Generated-program specificity and models.}
Some programs use benchmark-specific query templates and deterministic branches for recognizable problem structures.
Held-out evaluation tests new instances, not new domains or problem structures.
Both actors share the GPT-5.5 reflection endpoint; results may depend on the executor, tools, and reflection model.
Profile choices alone do not establish lower execution latency or deployment cost.

\section{Comparison with related optimization methods}
\label{app:related-comparison}

\paragraph{The editable unit determines what search can change.}
Table~\ref{tab:related-search-spaces} compares the editable components and inheritance rules of three close methods with \agepa.
Our focus is correspondence between specialists with evolving responsibilities and implementations.

\begin{table}[!htb]
\centering
\caption{\textbf{Related methods expose different units for optimization.} Each search object determines the editable components and how alternatives are selected or combined.}
\label{tab:related-search-spaces}
\small
\setlength{\tabcolsep}{4pt}
\renewcommand{\arraystretch}{1.12}
\begin{tabular}{@{}lll@{}}
\toprule
\parbox[t]{0.19\linewidth}{\raggedright Method} & \parbox[t]{0.34\linewidth}{\raggedright Editable unit} & \parbox[t]{0.42\linewidth}{\raggedright Selection or inheritance} \\
\midrule
\parbox[t]{0.19\linewidth}{\raggedright AutoPDL\\\citep{autopdl2025}} & \parbox[t]{0.34\linewidth}{\raggedright Prompting patterns, instructions, and demonstrations assembled into executable PDL programs.} & \parbox[t]{0.42\linewidth}{\raggedright Successive halving selects configurations within a predefined search space.} \\[5pt]
\parbox[t]{0.19\linewidth}{\raggedright JTPRO\\\citep{jtpro2026}} & \parbox[t]{0.34\linewidth}{\raggedright Global instructions and tool schemas, including argument descriptions.} & \parbox[t]{0.42\linewidth}{\raggedright Merges edits with the best context while maintaining correspondence through tool identity.} \\[5pt]
\parbox[t]{0.19\linewidth}{\raggedright Adaptive Auto-Harness\\\citep{adaptiveharness2026}} & \parbox[t]{0.34\linewidth}{\raggedright Harness branches with prompts, skills, tools, memory, and infrastructure for an open-ended task stream.} & \parbox[t]{0.42\linewidth}{\raggedright Git branches preserve alternative harnesses and their histories; a router selects a branch for a request.} \\[5pt]
\parbox[t]{0.19\linewidth}{\raggedright \agepa} & \parbox[t]{0.34\linewidth}{\raggedright A router and active description--program pairs, jointly evolved on a fixed request mixture.} & \parbox[t]{0.42\linewidth}{\raggedright Per-request selection retains alternatives; footprint alignment supplies correspondence for pair inheritance across candidates.} \\
\bottomrule
\end{tabular}
\end{table}

\paragraph{Diagnostic feedback is a separate design choice.}
Reflective Prompt Tuning uses diagnostic function calls and historical reports to revise a prompt \citep{rpt2026}.
VISTA separates labeled failure hypotheses from prompt rewriting and evaluates alternative hypotheses \citep{vista2026}.
Our two-stage reflection first selects a component to edit, then produces the edit.
This exposes routing and implementation changes as separate actions, but does not by itself verify a diagnosis or establish which action caused a score gain.
The recorded edit counts therefore describe search behavior rather than component contributions.

\paragraph{Relation to the broader search literature.}
Feedback-based edits and candidate combination also appear in the broader taxonomy of heuristic prompt optimization \citep{aposurvey2025}.
We retain reflection and an archive to jointly edit specialist responsibilities, routing, and executable programs.

\end{document}